\documentclass{elsarticle}
\usepackage{graphicx}
\usepackage{amssymb}
\usepackage{epstopdf}
\usepackage{bm}
\usepackage{color}

\def\0{{\boldsymbol 0}}

\def\q{{\bm q}}

\def\k{{\boldsymbol k}}

\def\p{{\boldsymbol p}}

\newcommand{\rmd}{{\rm d}}

\newcommand{\nn}{\nonumber\\ }
\newcommand{\beq}{\begin{eqnarray}}
\newcommand{\eeq}{\end{eqnarray}}

\journal{Physics Letters B}
\begin{document}

\begin{frontmatter}

\title{Massive renormalization scheme and perturbation theory at finite temperature}

\author{Jean-Paul Blaizot$^1$ and Nicol\'as Wschebor$^2$}

\address{$^1$Institut de Physique Th\'eorique, CNRS/URA2306, \\CEA-Saclay, 91191 Gif-sur-Yvette, France}
\address{$^2$Instituto de F\`{\i}sica, Faculdad de Ingenier\'{\i}a, 
Universidade de la Rep\'ublica, 11000 Montevideo, Uruguay}
\date{\today}

\begin{abstract}
\noindent
We argue that the choice of an appropriate, massive,  renormalization scheme can greatly improve the apparent convergence of perturbation theory at finite temperature. This is illustrated by the calculation of the pressure of a scalar field theory with quartic interactions, at 2-loop order. The result, almost identical to that obtained with more sophisticated resummation techniques,  shows a remarkable stability as the coupling constant grows, in sharp contrast with standard perturbation theory.  
\end{abstract}
\end{frontmatter}



\section{Introduction}

Over the last two decades considerable efforts have been devoted to understanding the behavior of perturbation theory in quantum field theory at finite temperature (for reviews, see \cite{Blaizot:2003tw,Kraemmer:2003gd,Andersen:2004fp}). These efforts are in part motivated by the physics of the quark-gluon plasma, and the  hope that QCD asymptotic freedom would allow for reliable calculations at sufficiently high temperature. It is well known, however,  that in QCD  infrared divergences inevitably occur and eventually cause a breakdown of QCD perturbation theory at a finite order. But even in theories where such a breakdown does not occur, such as in scalar theories, perturbation theory at finite temperature appears as poorly convergent as in QCD.

Two routes have been followed to try to overcome these difficulties. The first one is to include more and more terms into the perturbative series, hoping in doing so to improve the apparent convergence. Thus, the pressure of the massless scalar theory with a quartic interaction is now known up to order $g^8 \log (1/g)$  (see Ref.~\cite{Andersen:2009ct} and references therein). While definite improvements are observed at small coupling when high orders are taken into account, the bad behavior of the perturbative series resurfaces as soon as the coupling gets moderately large. 

The other route involves various  reorganizations of the perturbative expansion, such as screened perturbation theory \cite{Karsch:1997gj},  infinite resummations, the use of functional variational techniques such as the so-called 2PI (two particle irreducible) formalism \cite{Blaizot:2000fc}, or the  the non perturbative renormalization group (NPRG) \cite{Blaizot:2010ut}.  Remarkably, all these approaches produce results that remain stable as one increases the coupling constant, in sharp contrast with strict perturbation theory.  At the same time, some of these calculations suggest that the physics at moderate coupling is minimally non perturbative. In particular, calculations using the functional renormalization group  within the most sophisticated approximation scheme available \cite{Blaizot:2005xy,Blaizot:2010ut} yield results that do not  deviate much from simple self-consistent quasiparticle approximations (such as the lowest order 2PI approximation, or the local potential approximation of the NPRG \cite{Blaizot:2006rj}).

These latter results suggest to look for an underlying simplicity, and it is indeed  the purpose of this paper to report on progress in this direction. We shall argue that the difficulties encountered in finite temperature calculations  can be attributed to a large extent to inappropriate choices of renormalization schemes. The success of the NPRG invites us to look for a scheme where the thermal mass plays a central role, and also where a decoupling of modes occurs  when the typical scales exceed the temperature, both features that are automatically included in the NPRG. We shall exhibit such a massive renormalization scheme and show that it yields indeed a well behaved perturbative expansion. This will be illustrated in this letter with the calculation of the pressure of a scalar field theory.   More elaborate calculations will be presented in forthcoming publications. 

\section{Massive thermal scheme}\label{Sect:massicvescheme}

Let us start by recalling the origin of the difficulties with  standard perturbation theory at finite temperature, focusing on a scalar theory of massive  modes with quartic interactions $\sim g^2\varphi^4$ (see Eq.~(\ref{action}) below).  The expansion parameter, which is not simply $g^2$,  depends on the magnitude of the average fluctuations of the field, given by (we ignore here the vacuum fluctuations)
\beq\label{phi2}
\langle \varphi^2\rangle_\kappa\approx \int^\kappa \frac{d^3\p}{(2\pi)^3} \frac{n_\p}{E_\p}\approx T\kappa ,\qquad n_\p=\frac{1}{{\rm e}^{E_\p/T}-1},
\eeq
where  $E_\p=\sqrt{m^2+\p^2}$, with $m$ the mass,  $\kappa$ is an ultraviolet momentum cutoff and the approximation for $\langle \varphi^2\rangle_\kappa$ is valid for $\kappa\lesssim T$. Using this, we can define a dimensionless expansion parameter, $\gamma_\kappa$, as the ratio between the interaction energy ($\sim g^2 \langle \varphi^2\rangle_\kappa^2$), and the kinetic energy ($\sim\kappa^2 \langle \varphi^2\rangle_\kappa$) of modes with typical momentum $\kappa$, 
\beq
\gamma_\kappa\sim \frac{g^2\langle \varphi^2\rangle_\kappa}{\kappa^2}\sim \frac{g^2T}{\kappa}.
\eeq
Thus for $\kappa\sim T$, the expansion parameter is essentially the coupling constant $\gamma_T\sim g^2$. However, $\gamma_\kappa$ grows as $\kappa$ decreases. Eventually $\gamma_\kappa$ becomes of order unity when
$\kappa \sim g^2T$, at which point   standard perturbation theory breaks down. 

However, at least in the case of scalar theories, this scenario is too pessimistic, and this for two reasons. Observe first that when $m\ll\kappa \lesssim T$ the  theory behaves as
a massless three-dimensional  theory with (dimensionful) coupling $g^2T$.
 The associated {\it dimensionless} coupling can be identified to  $\gamma_\kappa$, and it obeys the 
 one-loop renormalization group equation (see Eq.~(\ref{3drunning}) below)
 \beq\label{RGgamma}
\kappa\frac{d \gamma_\kappa}{d\kappa}=-\gamma_\kappa + \frac{3}{16} \gamma_\kappa^2.
\eeq
The first term in this equation results from the analysis that we just presented, the second term is the one loop correction. This correction tames the growth of the coupling suggested by the first term, and indeed  the infrared fixed-point at $\gamma_*=16/3$ prevents the blow-up of $\gamma_\kappa$. The success of the expansion in $\epsilon=4-d$ indicates  that perturbation theory in the vicinity 
of this fixed point is reasonably accurate \cite{Amit:1984ms}.
The second reason which prevents the breakdown of perturbation theory is of course the generation of a thermal mass $m$ of order $gT$ which freezes  the running of the coupling   at the scale $\kappa\sim m$.\footnote{In QCD, the long wavelength ``magnetic'' fluctuations have a mass of order $g^2T$, which is not large enough to prevent the breakdown of perturbation theory.}

These considerations concerning the mechanisms that prevent the growth of the coupling,  make paradoxical the fact that standard perturbation theory behaves so badly at finite temperature. In fact, as we have already alluded to, the reason may not be perturbation theory itself, but rather the particular scheme used. Most studies are done in
non-decoupling schemes, such  as the $\overline{\rm MS}$ scheme, which is popular because of its technical simplicity.  But the discussion above suggests the use of a scheme where the matching between the four-dimensional and the three-dimensional
regimes when $\kappa\lesssim T$, as well as the suppression of fluctuations when $\kappa \lesssim m$,  are manifest order by order in perturbation theory. We shall now present such a scheme. \\

We consider the theory of a scalar field $\varphi$ with the action
\begin{equation}\label{action}
 S[\varphi]=\int_0^\beta d\tau\int d^dx \Big\lbrace \frac 1 2 \partial_\mu\varphi\partial_\mu\varphi+
 \frac{ {m_{\rm B}}^2 }{2} \varphi^2+\frac {g_{\rm B}^2} {4!} \varphi^4\Big\rbrace,
\end{equation}
where $m_{\rm B}$ and $g_{\rm B}$ denote the bare mass and coupling constant, respectively. 
The upper bound of the integration over the imaginary time $\tau$ is $\beta=1/T$, where $T$ is the temperature.  In line with the previous discussion, we introduce a specific renormalization scheme with the following, temperature dependent, renormalization conditions: 
\beq\label{renormcond}
 m^2&=\Gamma^{(2)}(\p=\0,\omega=0,T),\nonumber\\\
 1&=\frac{d\Gamma^{(2)}}{d  \p^2}(\p^2=\mu^2,\omega=0,T),\nonumber\\
 g^2&=\Gamma^{(4)}(\p^2_{sym}=\mu^2,\omega_i=0,T),
\eeq
where $\Gamma^{(2)}$ and $\Gamma^{(4)}$ are renormalized $n$-point functions, and $\p^2_{sym}$ refers to  a symmetric combination of 3-momenta.  There is of course a large flexibility in the choice of renormalization conditions. The scheme presented above satisfies the important requirement that the  renormalized coupling constant $g^2$ becomes independent of the temperature when $\mu \gg T$, so that we can isolate unambiguously thermal effects when comparing theories with different values of the coupling constant. The determination of the mass is more subtle. We want to use the thermal mass $m$ that enters the first renormalization condition (\ref{renormcond}) within the free propagators of the perturbative expansion. However, the renormalization condition does not determine $m$, it just fixes the finite parts of counterterms so that $m$ has a prescribed value. In order to relate this value to a mass that is known, we shall calculate $m_0$, the mass at $T=0$, at the order of perturbation theory at which we work. This will then provide a self-consistent equation (occasionally referred to as a gap equation) for the determination of $m$ as a function of $m_0$.  One may be worried about the fact that the present scheme involves counterterms whose finite parts depend on the temperature. However, this is not a problem if subdivergences are carefully eliminated, as they should.

\section{The 2-point function and the self-consistent thermal mass}

The one-loop contribution to the 2-point function is easily calculated: 
\begin{equation}\label{Gamma2oneloop}
 \Gamma^{(2)}(\p,\omega,T)=m^2+\delta m^2+\p^2
 + \frac {g^2T}{ 2} \sum_n \int \frac{\rmd^d q}{(2\pi)^d} \frac 1 {\omega_n^2+\q^2+m^2},
\end{equation}
where $\omega_n=2n\pi T$ is a Matsubara frequency, and we used the fact that the self-energy  is independent of  $\p^2$ in order to ignore the field renormalization factor. We have set $m_{\rm B}^2=m^2+\delta m^2$, where $m$ is the renormalized mass. Note that the renormalization of the coupling constant at one-loop order has an impact on the 2-point function only when this is calculated at 2-loop order (see next section). Accordingly, in Eq.~(\ref{Gamma2oneloop}),  $g$ is taken to be the  renormalized coupling constant. 
It is convenient to set
\beq
I(m)\equiv T \sum_n \int_\q \frac 1 {\omega_n^2+\q^2+m^2}=\int_\q \frac{1+2n_\q}{2E_\q}\equiv I_0(m)+I_T(m), 
\eeq
where we have introduced the shorthand notation for momentum space integrations, to be used throughout this paper: $\int_\q=\int \frac{\rmd^d q}{(2\pi)^d}$. In most of this paper, unless specified otherwise, we shall work in $d=3$ dimensions. In the formula above, $n_\q$ is the Bose statistical factor (see Eq.~(\ref{phi2})), and  $E_\q=\sqrt{\q^2+m^2}$. 

Specifying  $\p=0$ in Eq.~(\ref{Gamma2oneloop}), one gets:
\beq
\label{masserenorm}
 \Gamma^{(2)}(\p=0,\omega=0,T) =m^2+\delta m^2+\frac{g^2}{2} I(m).
\eeq
By imposing the first of the renormaliztion conditions~(\ref{renormcond}), namely $m^2=\Gamma^{(2)}(\p=0,\omega=0,T)$, one deduces the value of the counterterm
\beq\label{ctmass}
\delta m^2=-\frac{g^2}{2} I(m).
\eeq
At this point, the mass $m$ that enters the renormalization condition is in general unknown. However, it can be related to the renormalized mass at zero temperature, which we assume to be a known quantity.  To that aim, we repeat, at $T=0$,  the leading order perturbation theory calculation that yields Eq.~(\ref{masserenorm}). We get
\beq
\label{masserenorm0}
 \Gamma^{(2)}(\p=0,\omega=0,T=0) =m^2+\delta m^2+
  \frac{g^2}{ 2} I_0(m)=m^2-\frac{g^2}{2} I_T(m),\eeq
  where, in the last step, we have used the expression (\ref{ctmass}) of the mass counterterm. 
By identifying the left-hand-side of Eq.~(\ref{masserenorm0}) to $m_0^2$, with $m_0$ the $T=0$ renormalized mass, one gets the looked for relation between $m_0$ and $m$
\beq\label{gap0T}
m^2=m_0^2+\frac{g^2}{2}I_T(m).
\eeq
Eq.~(\ref{masserenorm0}), which expresses a zero temperature quantity in terms of finite temperature ones, is not usual. But in fact,  Eq.~(\ref{gap0T}) tells us that the thermal mass $m$ is adjusted so that  $m^2-\frac{g^2}{ 2} I_T(m^2)$ is independent of the temperature. \\

Eq.~(\ref{gap0T}) is a self-consistent equation for $m^2$.  This self-consistent determination of the thermal mass has already been recognized in other approaches as an important factor that stabilizes the weak coupling expansion,  and it constitutes an essential aspect of the present renormalization scheme. 
At this point it is instructive to contrast the result obtained from solving the self-consistent equation (\ref{gap0T}) with that given by standard perturbation theory. Recall that standard perturbative approaches use the zero temperature quantities as their starting point, and consider finite temperature effects as corrections. In the present context, this is equivalent to solve Eq.~(\ref{gap0T}) by expanding with respect to the $T=0$ parameters, i.e., treating $m-m_0$ as a perturbation. To perform this expansion, we assume a small  coupling constant $g$, and a vanishing  zero temperature mass, $m_0=0$.
The leading order contribution is then obtained by neglecting the mass in the calculation of $I_T$.  Using $I_T(m=0)=T^2/12$ in Eq.~(\ref{gap0T}), one gets $m^2=g^2T^2/24\equiv \hat m^2$, where $\hat m$ is the standard Hard Thermal Loop (HTL) result (see e.g. Ref.~\cite{Blaizot:2003tw}). One can go beyond this leading order by treating the mass as a small correction in the calculation of $I_T$, that is we define
\beq
m^2=\hat m^2+\Delta m^2,\qquad \Delta m^2=\frac{g^2}{2} \left( I_T(\hat m)-I_T(0) \right).
\eeq
The calculation of $\Delta m^2$ is complicated by the fact that a naive expansion in powers of $\hat m$ generates infrared divergences. This is because the integral that yields  $\Delta m^2$  is no longer dominated by momenta of order $T$, but rather by momenta of order  $\hat m \sim gT\ll T$. To obtain  the limit 
$\hat m \ll T$ we first perform the change of variables $q=\hat m y$, and only then expand in powers of  $\hat m$. One then gets:
\beq
\label{correcmth2}
 \Delta m^2&=&  \frac{g^2\, {\hat m}^{2}}{4\pi^{2}} \int_0^\infty \rmd y \,y^2
  \left(\frac{1}{\sqrt{y^2+1}(e^{\beta \hat m \sqrt{y^2+1}}-1)} -\frac{1}{y (e^{\beta \hat m y}-1)} \right) \nonumber\\
  &\approx & - \frac{g^2\,T\, \hat m}{4\pi^2} \int_0^\infty \,\rmd y\,\frac{1}{y^2+1} 
  =-\frac {g^3 T^2}{16\pi\sqrt{6}},
  \eeq
which is the known result, usually derived from HTL resummations. We note that the $g^3$  correction is negative, implying that $m^2$ eventually becomes negative as the coupling grows, signaling a breakdown of perturbation theory for $g^2/24\sim 1$ (see Fig.~\ref{courbem} below). By contrast, the solution of the gap equation remains perfectly regular as the coupling increases.

\section{The 4-point function and the $\beta$ function}

We turn now to the calculation of the 4-point function at order one-loop. Since the running of the coupling does not play an essential role in the leading order calculation of either the mass (previous section) or the pressure (next section), our goal here is merely to illustrate generic features of the $\beta$-function in the present massive scheme,  in particular the role of the mass in cutting off the flow, as well as the decoupling property mentioned in Sect~\ref{Sect:massicvescheme}, i.e., the transition form 4-dimensional to 3-dimensional running when the scale passes below $2\pi T$.

In order to implement the renormalization condition (the last condition (\ref{renormcond})) we need the 4-point function for vanishing external frequencies and a symmetric configuration of momenta (here 
$\p^2=(\p_1+\p_2)^2=(\p_1+\p_3)^2=(\p_2+\p_3)^2$):
\beq
\Gamma^{(4)}(\p^2,\omega_i=0,T)&\!=\!& \!g^2_B- \frac{3 g^4}{2} T\sum_n \int_\q \frac 1 {\omega_n^2+\q^2+m^2}
\frac 1 {\omega_n^2+(\q+\p)^2+m^2}.\nn
 \eeq
(We do not need to worry about the potential logarithmic divergence of the integral  since this will drop in the differentiation with respect to $\mu$.) 
Recalling that $g^2(T,\mu)= \Gamma^{(4)}(\p^2_{sym}=\mu^2,\omega_i=0,T)$, and that $g_B$ does not depend on $\mu$, one gets
\beq\label{betafunction}
\beta(g^2,\mu,m,T)&\equiv& \mu \frac{\rmd g^2(T,\mu)}{\rmd \mu}\nn
&=& 6 g^4 T \sum_n \int_0^1 dx \int_\q \frac {x(1-x) \mu^2} {(\omega_n^2+\q^2+m^2+x(1-x) \mu^2)^3},\nn
 \eeq
where only the explicit $\mu$ dependence was involved in the differentiation, i.e., we have ignored terms of order $g^6$ or higher. 
By performing (for  $d=3$) the integrals over the  Feynman parameter and the momentum, one can put the $\beta$-function in the form
 \beq\label{betafunction1}
\beta(g^2,\mu,m^2,T)
=\beta_0(g^2) \, F(\bar\mu,\bar m^2),\qquad \bar\mu\equiv\frac{\mu}{2\pi T},\quad \bar m^2\equiv \frac{m^2}{(2\pi T)^2}, 
\eeq
where  $F(\bar\mu,\bar m^2)$ is the dimensionless function
\beq\label{defF}
F(\bar\mu,\bar m^2)= \sum_n \Bigg\lbrace
-\frac{2 \sqrt{n^2+\bar m^2}}{\bar\mu^2+4 (n^2+\bar m^2)}+\frac{1}{\bar\mu} \mathrm{Arccot}\Big(2 \sqrt{\frac{n^2+\bar m^2}{\bar\mu^2}}\Big)
\Bigg\rbrace,
 \eeq
 and
 \beq\label{beta0}
 \beta_0(g^2)=\frac{3g^4}{16\pi^2}
 \eeq
  is the  standard one-loop $\beta$-function for a scalar field theory in 3+1 dimensions, in a mass independent scheme.   
\\

The sum over the Matsubara frequencies in the expression (\ref{defF})  is dominated by values of $n$ such that $\sqrt{n^2+\bar m^2}\lesssim \bar\mu$. When  $\mu \gg 2\pi T$  ($\bar\mu\gg 1$), one expects to recover the known result for the zero temperature theory in 4 dimensions. This is indeed the case:  In this regime, one can ignore the discrete nature of the sum over  Matsubara
frequencies, and replace the sum by an integral. The function $F(\bar\mu,\bar m^2)$  becomes then a function of $x\equiv m^2/\mu^2$ only, 
  \begin{equation}
F_0(x)=
1+\frac{2 x}{\sqrt{1+4 x}} \log\Big(\frac{\sqrt{1+4 x}-1}{\sqrt{1+4 x}+1}\Big)
,\qquad x\equiv \frac{m^2}{\mu^2}.
\end{equation}
When $\mu^2\gg m^2$, $F_0(x)\to 1$,  and $\beta(g^2,\mu,m^2,T)\to 
\beta_0(g^2)$. 
 In the opposite regime where
 $\mu^2 \ll m^2$, the $\beta$ function is suppressed,
  \begin{equation}
\beta(g^2,\bar\mu,\bar m^2)
\simeq \beta_0(g^2)\,
\frac{\mu^2}{6 m^2}.
\end{equation}
In this regime, the flow eventually stops, the fluctuations at a scale smaller than the mass playing essentially no role.\\

In the regime of high temperature, $\mu\ll 2\pi T$, one may restrict the sum over the Matsubara frequencies to the lowest one, i.e., to $n=0$. The function in Eq.~(\ref{defF}) becomes then
\beq
F_T(\bar\mu,\bar m^2)= \frac{1}{\bar\mu}\left( -\frac{2\sqrt{x}}{1+4x}+\mathrm{Arccot}(2\sqrt{x}\,) \right), \qquad x\equiv \frac{m^2}{\mu^2}.
\eeq
When $\mu\gg m$, the $\beta$-function reduces to 
\beq
\beta(g^2,\bar\mu,\bar m^2)\simeq\beta_0(g^2) \, \frac{\pi^2 T}{\mu}, 
\eeq
and the beta function is enhanced as soon as $\mu\lesssim 2\pi T$. In fact, in this high temperature regime, the theory becomes essentially 3-dimensional, with a dimensional coupling constant $g^2T$. One can introduce (for $\mu\gg m$) an effective dimensionless coupling $\gamma=g^2T/\mu$, whose  flow  is easily obtained:
 \beq\label{3drunning}
\mu\partial_\mu \gamma=-\gamma+\frac{3 \gamma^2}{16 \pi} \,f_T(x), \qquad x=\frac{m^2}{\mu^2},
\eeq
with $f_T(x)\equiv 2\bar\mu\, F_T(\bar\mu,\bar m^2)$. The function $f_T(x)$ is a steeply decreasing function of $x$, starting at $f_T(0)=\pi$, so that in the regime $\mu\gg m$, one recovers the $\beta$-function (\ref{RGgamma}) mentioned in Sect.~\ref{Sect:massicvescheme}.
 When $\mu\lesssim m$,   the flow of the coupling is  suppressed.  We have indeed, for $\mu\ll m$, 
 \beq
 \beta(g^2,\bar\mu,\bar m^2)\simeq\beta_0(g^2) \, \frac{\bar\mu^2}{12\bar m^3}.
\eeq
\\

\begin{figure}[h]
\begin{center}
\hspace{-2em}\includegraphics[scale=0.48,angle=0]{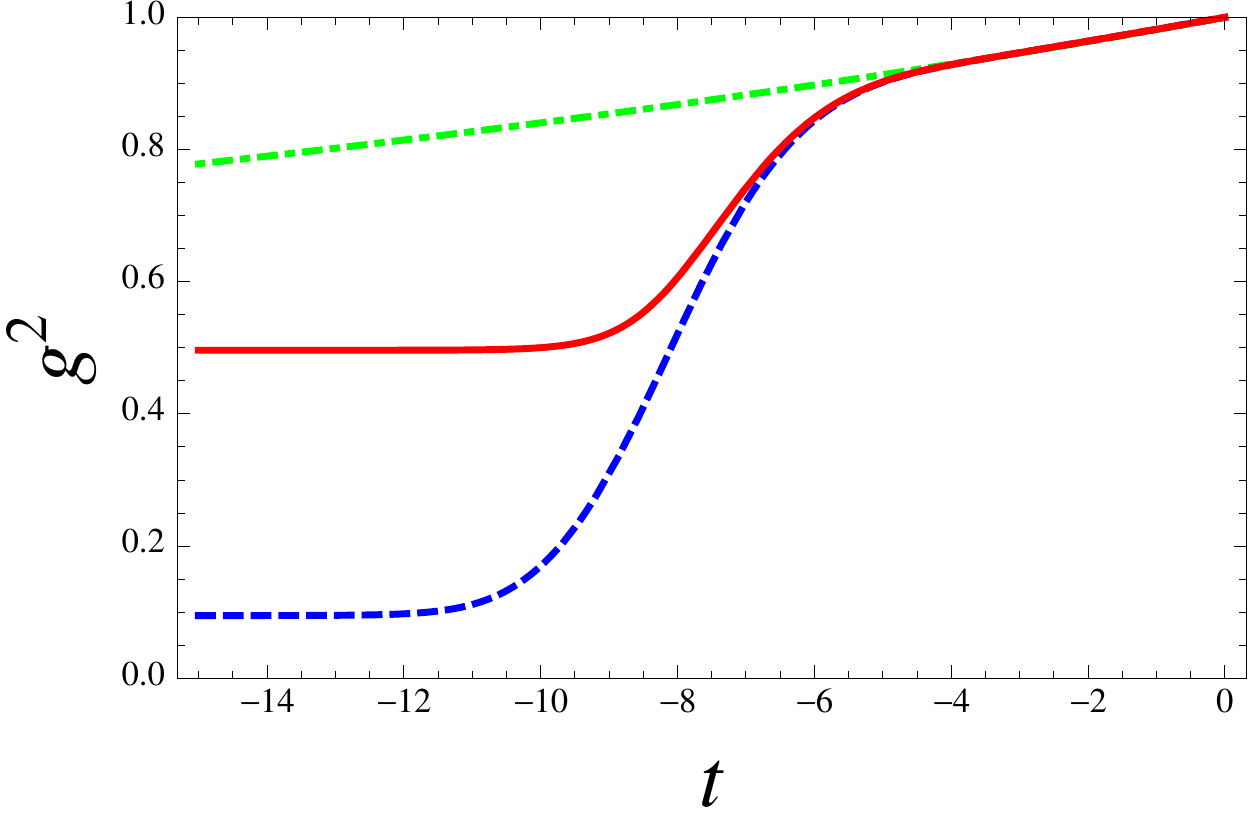}\includegraphics[scale=0.48,angle=0]{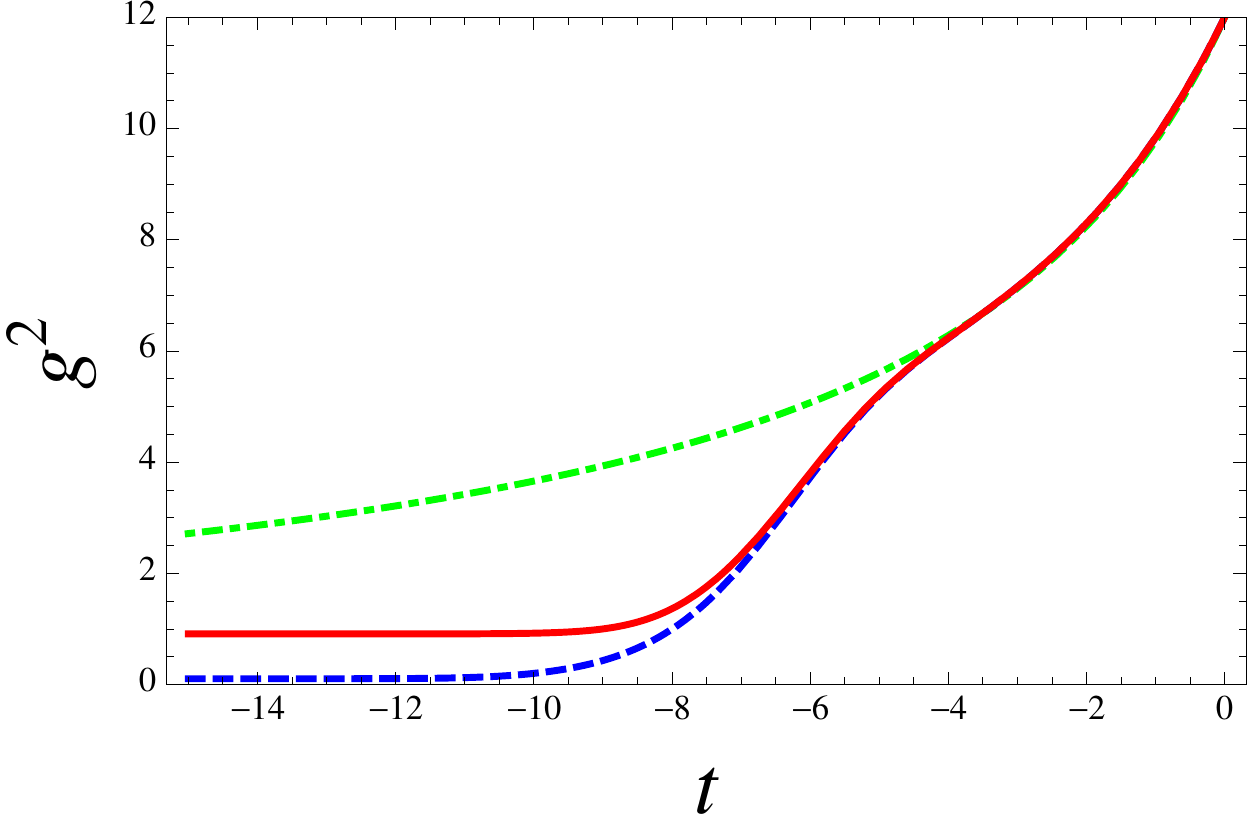}
\caption{(Color online.) The running coupling $g^2$ obtained by integrating the $\beta$-function, as a function of $t=\ln(\mu/\Lambda)$. The initial coupling, at the scale $\Lambda/(2\pi T)=10^2$, is $g^2=1$ (left) and $g^2=12$ (right). The dot-dashed line (green) is obtained by integrating the standard $T=0$ $\beta$-function, Eq.~(\ref{beta0}). The dashed (blue) line is obtained by integrating the full $\beta$-function with a (fixed) mass $\bar m=m/(2\pi T)=10^{-3}$.  The full line is obtained similarly for $\bar m=10^{-2}$. The scale $\bar \mu=1$ ($\mu=2\pi T$) corresponds to $t=-4.6$,  $\bar\mu=10^{-3}$  to $t=-11.5$, and $\bar\mu=10^{-3}$ to $t=-9.2$. \label{fig:betaT}
}
\end{center}
\end{figure}

All the features that we have discussed in this section are illustrated in Fig.~\ref{fig:betaT}, for small ($g^2=1$) and moderate ($g^2=12$) coupling. The acceleration of the flow as $\mu$ gets lower than $2\pi T$ is clearly seen, as well as the saturation of the flow when $\mu\lesssim m$. These features, generic of the present massive scheme, are reminiscent of those observed in the non perturbative renormalization group approach (see e.g. \cite{Blaizot:2010ut,Blaizot:2006rj}), and indeed, as we have already emphasized,  the main properties of the present renormalization scheme are automatically included in such an approach. Note that the calculations presented  in Fig.~\ref{fig:betaT} are done with a fixed value of the mass. In an actual calculation, the self-consistent thermal mass should be used, but that would not alter the general behavior illustrated in Fig.~\ref{fig:betaT}.

\begin{figure}
\begin{center}\hspace{-2em}\includegraphics[scale=0.23,angle=-90]{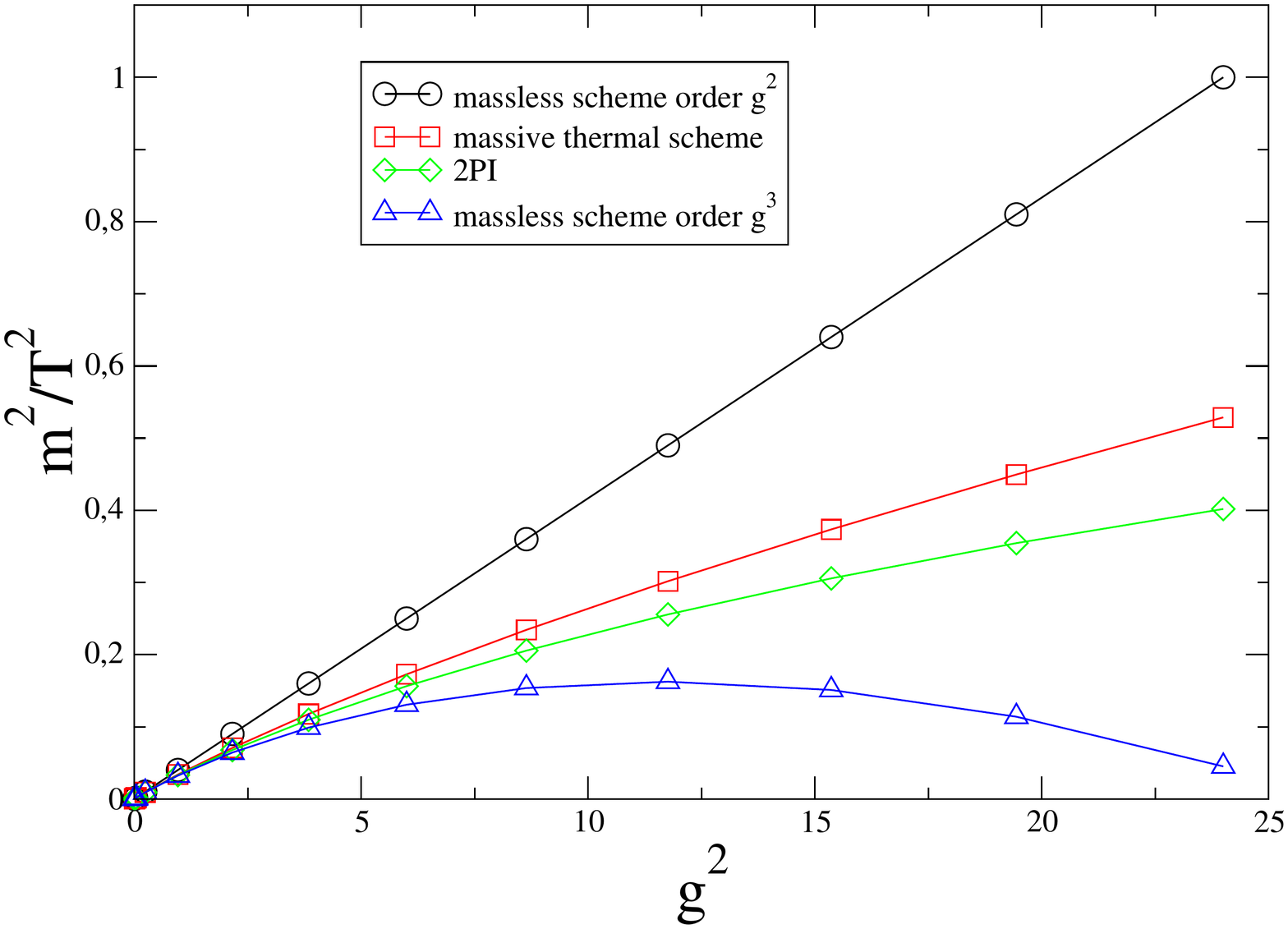}\includegraphics[scale=0.23,angle=-90]{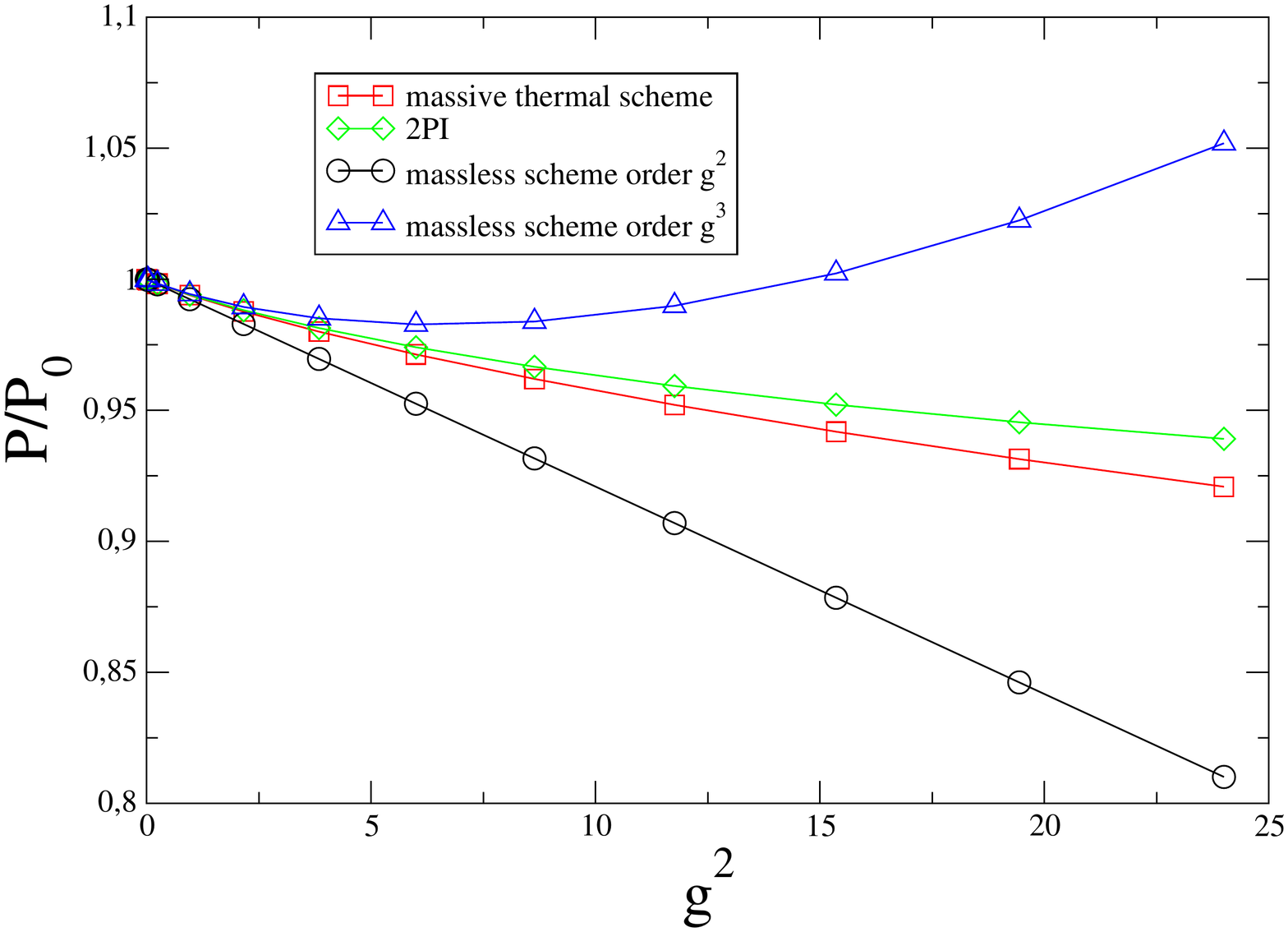}
\caption{(Color online.) The quantities $m^2/T^2$  (left) and $P/P_0$ (right) as a function of the (fixed) coupling constant. The mass $m$ is obtained as the solution of the gap equation (\ref{gap0T}),  for $m_0=0$ (red, squares). The pressure $P$ (with $P_0$ the non interacting pressure) is obtained from Eq.~(\ref{pressureT}) with $m$ solution of the gap equation (red, squares).  The black  curves (circles) are the leading results of order $g^2$, while the blue curves (triangles) includes the $g^3$ correction. The green curves (diamond) display, for comparison, the results of the 2PI approximation, with the coupling constant  $g^2=g^2(\mu=2\pi T)$. \label{courbem}
 }
\end{center}
\end{figure}

 \section{Calculation of the pressure}
 
 We turn now to the calculation of the pressure, and we limit ourselves to the calculation of the  leading order, 2-loop, perturbative correction. We have 
\beq\label{pressure1a}
 P(T)=-\frac T 2 \sum_n \int_\q \log (\omega_n^2+\q^2+m^2)-\frac{g^2}{8} I(m^2)^2 - \frac{1}{2} \delta m^2 I(m)-\delta P, 
\eeq
where the coupling constant is evaluated at $\mu=0$ (there is no momentum dependence in the vertex).
In the expression above,  $\delta P$ is a counterterm that absorbs the overall divergence that remains when mass subdivergences (the next to last term) have been eliminated, and whose finite part is adjusted so that $P(T=0)=0$. In the foregoing calculation, the following expression is useful\footnote{The  summation over the Matsubara frequencies requires a regularization that eliminates an infinite constant that does not depend on $T$ or $m$. Such a regularization is implied in writing Eq.~(\ref{determinant})}
\beq\label{determinant}
\frac{T}{2}\sum_n \ln(\omega_n^2+\omega_\q^2)=\frac{1}{2}\omega_\q+T\ln \left(1-{\rm e}^{-\omega_\q/T}  \right).
\eeq
Using this relation we rewrite Eq.~(\ref{pressure1a}) as 
\beq
 P(T)=-\frac{1}{2}\int_\q\omega_\q^T-T\int_\q\ln \left(1-{\rm e}^{-\omega_\q^T/T}  \right)-\frac{g^2}{ 8} [I(m)]^2 - \frac{1}{2} \delta m^2 I(m)-\delta P, \nn
\eeq
where $\omega_\q^{T}=\sqrt{\q^2+m^2}$.
By using the expression (\ref{ctmass})  of the mass counterterm, one easily gets
\beq
P(T)=-\frac{1}{2}\int_\k \omega_\k^{T}-T\int_\q\ln \left(1-{\rm e}^{-\omega_\q^T/T}  \right) +\frac{g^2}{8}[I(m^2)]^2-\delta P.
\eeq
This expression is still divergent, but the divergence is a global divergence present in the $T=0$ contribution. Let us then calculate the zero temperature pressure, in leading order perturbation theory. We obtain
\beq\label{pressure0}
 P(T=0)&=&-\frac{1}{2}\int_\q\omega_\q^T-\frac{ g^2}{ 8} [I_0(m)]^2 - \frac{1}{2} \delta m^2 I_0(m)-\delta P, \nn
 &=& -\frac{1}{2}\int_\q\omega_\q^T+\frac{g^2}{ 8}[ I_0(m)]^2 +\frac{g^2}{4} I_T( m) I_0(m)-\delta P.
\eeq
Subtracting this from $P(T)$ leaves us with the manifestly finite expression
\beq\label{pressureT}
 P(T)-P(T=0)=-T\int_\q\ln \left(1-{\rm e}^{-\omega_\q^T/T}  \right)+\frac{g^2}{ 8} [I_T(m)]^2. 
\eeq

As was the case of Eq.~(\ref{masserenorm0}), the expression  (\ref{pressure0}) has the unusual feature of  giving the zero temperature pressure  in terms of temperature dependent quantities. However, it is easy to show that $P(T=0)$ is independent of the temperature, as it should. Indeed, 
note that
\beq\label{expandomega}
\omega_\q^T=\sqrt{\q^2+m^2}\simeq \omega_\q^{T=0}+\frac{m^2-m_0^2}{2\omega_\q^{T=0}},\qquad  \omega_\q^{T=0}=\sqrt{\q^2+m_0^2}.
\eeq
Then by using Eq.~(\ref{gap0T}) for $m^2-m_0^2$, one gets, in leading order in $g^2$, 
\beq\label{P0ct}
 P(T=0)= -\frac{1}{2}\int_\q\omega_\q^{T=0}+\frac{g^2}{8} [I_0(m_0)]^2 -\delta P,
\eeq
which is  independent of the temperature. Note that in arriving at this result, the self-consistent relation (\ref{gap0T}) for $m$ has played an essential role. 
The  equation (\ref{P0ct})  can be used to fix the counterterm $\delta P$ (by requiring  that $P(T=0)=0$).

\begin{figure}
\begin{center}
\includegraphics[scale=0.35,angle=0]{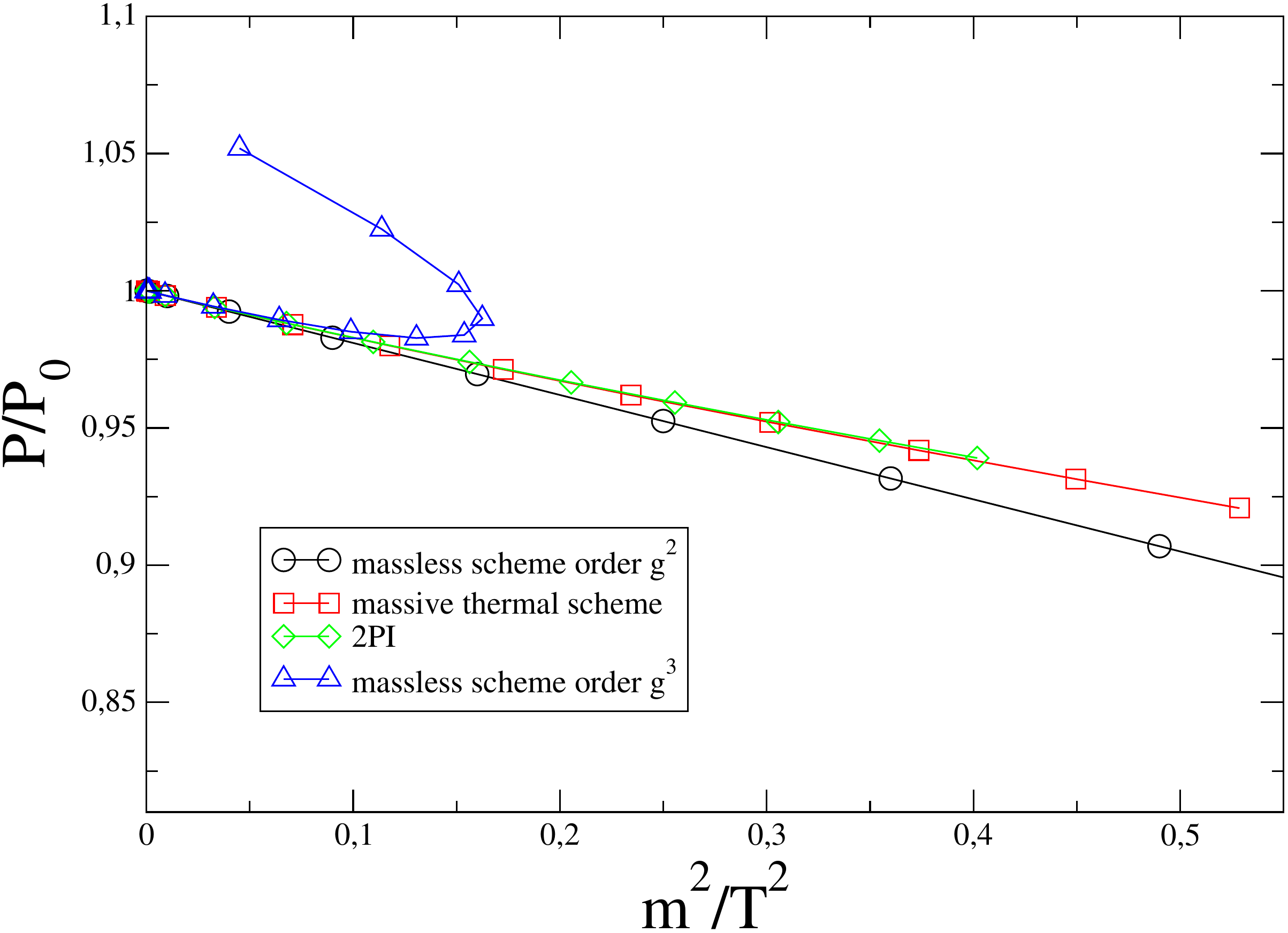}
\caption{(Color online.) The pressure $P/P_0$ (with $P_0=\pi^2T^4/90$ the free pressure) versus  $m^2/T^2$.\label{pvm}}
\end{center}
\end{figure}

As we have done for the mass, it is a simple exercise to show that the expression (\ref{pressureT}) for the pressure reproduces the first terms, i.e. the $g^2$ and $g^3$ contributions of the standard weak coupling expansion
\cite{Blaizot:2000fc}
\beq\label{pressurepert}
P(T)-P(T=0)\approx \frac{\pi^2 T^4}{90}-\frac{\hat m^2}{48}T^2+\frac{\hat m^3 T}{12\pi},\qquad \hat m^2=\frac{g^2}{24} T^2.
\eeq
The large differences between successive orders that was observed for the mass is also visible for the pressure. However, as illustrated in Fig.~(\ref{courbem}), the full result is much more stable as the coupling increases. Note that the curves in Fig.~(\ref{courbem}) correspond to calculations at  fixed values of the coupling constant.

The curves in Fig.~(\ref{courbem}) contain a comparison with the  results of the  2PI approximation, which, at 2-loop order yields for the pressure \cite{Blaizot:2000fc}
\beq
P=-T\int_\q\ln \left(1-{\rm e}^{-\omega_\q^T/T}\right)+\frac{m^2}{4}I_T+\frac{m^4}{128\pi^2},
\eeq
with $m$ solution of the gap equation
\beq\label{gap2PI}
m^2 =\frac{g^2}{2} I_T+\frac{g^2 m^2}{32\pi^2}\left( \ln\frac{m^2}{\mu^2}-1 \right).
\eeq
The latter expression is obtained by using dimensional regularization and minimal subtraction \cite{Blaizot:2000fc}, and $g=g(\mu)$ is the running coupling. The natural choice of scale in this scheme is  $\mu=2\pi T$, and this is the value of the fixed coupling that is used in drawing the curves in  Fig.~(\ref{courbem}).\footnote{Note that at this level of approximation, 
the $\beta$-function is $\mu{\partial g^2}/{\partial\mu}={g^4}/{(16\pi^2)}$, i.e.,  one third of the perturbative beta function (\ref{beta0}). This is because only one channel out of three is taken into account at this level of the 2PI approximations. However  the running of the coupling plays essentially no role in the present calculation.}  The results of the 2PI approximation are quite similar to those of the present scheme. We note again the stabilizing effect of the mass resummation, that is accounted for here by an equation quite similar to that of the massive scheme (cp. Eqs.~(\ref{gap0T}) and (\ref{gap2PI})). 

The relations between $m^2$ or $P$ and  the coupling constant depend of course on the renormalization scheme. To make a fair comparison between the two approximations, and eliminate most of this scheme dependence, we have plotted  in Fig.~(\ref{pvm}) the pressure as a function of the thermal mass. Both quantities are  determined as a function of the coupling as indicated earlier. However, the 2PI result is actually independent of the choice of the renormalization scale $\mu$. As for the massive scheme, it does not matter whether $m$ and $P$ are calculated with the coupling $g(2\pi T)$ or the coupling deduced from that particular value by letting the scale run down from $2\pi T$ to zero. Figure~(\ref{pvm})  shows that the 2PI result is almost identical to that of the massive scheme, and both are quite distinct from that of standard perturbation theory.

\section{Conclusions}

We have shown that, by adopting a massive, temperature dependent, renormalization scheme, one can greatly improve the apparent convergence of perturbation theory at finite temperature. The thermal mass scheme presented in this letter yields a well behaved perturbative expansion in powers of $g^2$. The familiar non analytic contributions (such as the contribution of order $g^3$) are hidden in the dependence of the mass on the coupling constant (and would resurface if we were to expand in powers of $m$, which we do not need to do of course). Calculations in a massive scheme are more difficult than, say, standard calculations for massless theories using the $\overline{\rm MS}$ scheme.  But this is a small price to pay if accurate results are obtained with the first few orders of perturbation theory. An additional complication of the massive scheme is that a self-consistent equation  for the mass needs to be solved.  Although it looks very similar to that of the 2PI scheme,  this equation is in fact much simpler: it is an equation for the mass, instead of an equation for the fully self-consistent propagator. 

For illustration purpose, we have presented  a comparison with the 2PI approximation. But we should mention here that various methods to reorganize perturbation theory have been proposed over the years. Some bear strong similarity with what we propose here, as they emphasize the need  to expand around a massive theory. Among those, we have already  mentioned screened perturbation theory \cite{Karsch:1997gj,Andersen:2000yj,Andersen:2001ez}, where one expands with respect to an auxiliary  mass whose value  is eventually determined by various prescriptions. We should also mentioned numerous versions of optimized perturbation theory,  where the variation of the auxiliary mass is parametrically correlated to that of the coupling constant  (see for instance \cite{Kneur:2004yb}, and references therein). We believe that the present scheme is both conceptually and technically simpler. Applications of this scheme to high  order calculations and more sophisticated field theories will be presented in forthcoming publications.

\section*{Acknowledgements}
This work is 
supported by the European Research Council under the
Advanced Investigator Grant ERC-AD-267258. NW acknowledges the  support of the ECOS Uruguay-France cooperation program. He also would like to thank the hospitality of the Institut de Physique Th\'eorique of CEA Saclay, where much of this work has been done.

\bibliographystyle{unsrt}

\end{document}